\documentclass[fleqn,usenatbib]{mnras}

\usepackage[T1]{fontenc}
\usepackage{ae,aecompl}

\usepackage{graphicx}	
\usepackage{amssymb}	
\usepackage{longtable}
\usepackage{pdflscape}
\usepackage{lscape}
\usepackage{booktabs}
\usepackage{csquotes}
\usepackage{color, colortbl}
\usepackage{amstext}
\usepackage{wasysym}
\usepackage{multirow}
\usepackage{natbib}
\usepackage{ulem}

\def\degr{\hbox{$^\circ$}}

\usepackage{t1enc}
\usepackage{times}
\usepackage[utf8]{inputenc}
\usepackage{graphicx}
\usepackage{footnote}
\usepackage{pifont}
\usepackage{lscape}
\usepackage{booktabs}
\usepackage{csquotes}
\definecolor{LightGray}{gray}{0.9}
\definecolor{LightGray1}{gray}{0.8}
\definecolor{pad}{rgb}{0.77,0.07,0.77}

\def\degr{\hbox{$^\circ$}}

\title[Statistics of long-period comets]{Discovery statistics and the 1/a-distribution of long-period comets detected over the 1801--2017 period.}

\author[M. Kr\'{o}likowska \& P. A. Dybczy\'{n}ski]{Ma{\l}gorzata Kr\'{o}likowska ,$^{1}$\thanks{E-mail: mkr@cbk.waw.pl}
Piotr A. Dybczy\'{n}ski,$^{2}$\thanks{E-mail: dybol@amu.edu.pl}
\\
$^{1}$Space Research Centre of the Polish Academy of Sciences, Bartycka 18A, 00-716 Warsaw, Poland\\
$^{2}$Astronomical Observatory Institute, Faculty of Physics, A.~Mickiewicz University, S{\l}oneczna 36, 60-286 Pozna\'{n}, Poland
}

\date{Accepted XXX. Received YYY; in original form ZZZ}

\pubyear{2018}

\begin{document}
\label{firstpage}
\pagerange{\pageref{firstpage}--\pageref{lastpage}}
\maketitle




\begin{abstract}
For the last two decades we have been observing a huge increase in discoveries of long-period comets (LPCs), especially those with large-perihelion distances.

We collected data for a full sample of LPCs discovered over the 1801--2017 period including their osculating orbits, discovery moments (to study the discovery distances), and original semimajor axes (to study the number ratio of large-perihelion to small-perihelion LPCs in function of $1/a$-original, and to construct the precise distribution of an $1/a$-original). To minimize the influence of parabolic comets on these distributions we determined definitive orbits (which include eccentricities) for more than 20\,LPCs previously classified as parabolic comets.

We show that the percentage of large-perihelion comets is significantly higher within Oort spike comets than in a group of LPCs with $a<10\,000$\,au, and this ratio of large-perihelion to small-perihelion comets for both groups has grown systematically since 1970. The different shape of the Oort spike for small-perihelion and large-perihelion LPCs is also discussed. A spectacular decrease of the ratio of large-perihelion to small-perihelion LPCs with the shortening of semimajor axis within the range of 5000--100\,au is also noticed.

Analysing discovery circumstances, we found that Oort spike comets are discovered statistically at larger geocentric and heliocentric distances than the remaining LPCs. This difference in the percentage of large-perihelion comets in both groups of LPCs can probably be a direct consequence of a well-known comets fading process due to ageing of their surface during the consecutive perihelion passages and/or reflects the different actual $q$-distributions.

\end{abstract}




\section{Introduction}\label{sec:intr}

In this century the spectacular increase in discoveries of long-period comets (orbital period greater than 200\,yr,  $a > \sim34$ au, hereafter LPCs) is observed. At the same time, there is no statistical comparison of these findings, for example in the context of the previous history of discoveries. Thus, we present such an analysis for all LPCs discovered since 1800 as well as for various statistical comparisons using some subsamples of LPCs. We focus mainly on differences between Oort spike comets and LPCs with more tight orbits, and differences between small- and large-perihelion LPCs\footnote{We use concise terms: {\it large-perihelion comets} and {\it small-perihelion comets} for objects moving on orbits with a perihelion distance over 3.1\,au or below 3.1\,au, respectively.} in the context of their discovery statistics and $1/a_{\rm ori}$-distributions. 

Several valuable papers and books containing a review of a LPCs population were published so far. To mention some of them we should start with \cite{vanWoerkom:1948}, through \cite{wiegert-tre:1999}, \cite{Francis:2005}, \cite{Rickman:2014}, \cite{FRFV:2017} and end with two excellent books on this subject: \cite{fernandez_book:2005} and \cite{Rickman-book:2018}. In comparison with these previously published results our paper benefits from more numerous and more precise LPCs orbits. 

In the present paper we use the term 'Oort spike' in two slightly different meanings. Strictly speaking, if we treat the Oort spike comets as coming from the hypothetical Oort cloud \citep{oort:1950} we have to restrict ourselves to elliptical original orbits. Therefore the strict understanding of this term corresponds to a definition based on the semimajor axis reciprocal of the original comet orbit: $0 < 1/a_{\rm ori} < 0.0001$au$^{-1}$.

But in many cases it is more appropriate to use this term in a wider sense. Apart from observational uncertainties the main reason for including some comets outside the above interval of $1/a_{\rm ori}$ to the Oort spike comets, especially those with small perihelion distances, is the existence of the non-gravitational forces (hereafter NGF). Except perhaps a few special cases we believe that probably all LPCs with negative  $1/a_{\rm ori}$ are the result of omitting or inadequate treating of NGF during the definitive orbit determination process. This widely accepted opinion is strongly upheld by numerous cases, when  NGF parameters are successfully determined for LPCs, see for example  \citet{marsden-sek-ye:1973,krolikowska:2001,krolikowska:2004,krolikowska:2006a,krolikowska:2014,kroli-dyb:2010}. The typically observed effect of shifting toward positive values of $1/a_{\rm ori}$ after including NGF into the dynamical model convince us to classify as the Oort spike comets also those, where NGF are indeterminable. 

Additionally, in this wider sense,  we also treat as potentially belonging to the Oort spike all LPCs with $1/a_{\rm ori}$ slightly greater than 0.0001~au$^{-1}$, just accounting for the observational uncertainties. Such a wider understanding of the 'Oort spike' which includes both negative and positive 'wings' of the spike will be clearly notified in the remaining text.

In this paper we present various statistics for over one thousand of LPCs and try to interpret some interesting differences in distributions between Oort spike and the remaining LPCs (sections~\ref{sec:stat}--\ref{sec:perihelion}) and next we study $1/a_{\rm ori}$ distribution for over 880 of them (sections~\ref{sec:a_original} and \ref{sec:Oortspike}). To face this last task, we collected a list of original $1/a$ for the whole sample of LPCs with definitive orbits basing on our previous orbital determinations \citep[see for example][KD17 and KD13, respectively]{kroli_dyb:2017,kroli-dyb:2013}, the printed version of Catalogue of Cometary Orbits \citep{marsden-cat:2008} and Web-sources, mainly IAU Minor Planet Center Database Search Engine\footnote{see {https://www.minorplanetcenter.net/db\_search/}} and Nakano Notes\footnote{see {http://www.oaa.gr.jp/{\textasciitilde}oaacs/nk.htm}}. In the last section we summarize our findings.

For all LPCs the semimajor axis is the most difficult  orbit element to determine. It is a common practice to assume a parabolic orbit in cases when it is too difficult to obtain the orbital eccentricity. In the studied sample of LPCs we found such a situation for almost 20\,per cent of objects. The indeterminacy of $1/a$ forced us to exclude them from some part of this study, especially in Sections~\ref{sec:a_original} and~\ref{sec:Oortspike}. But parabolic orbits contain valuable approximate information on the remaining orbital elements of LPCs, so we carefully examined the possibility of including these comets into various statistics, see Sections~\ref{sec:stat_large_small}--\ref{sec:a_original}. We also attempted to determine several definitive orbits for many  objects previously classified as 'parabolic comet' to minimize the unknown influence of parabolic comets on LPCs statistics (see Section~\ref{sec:stat_par}).

\section{Discovery statistics for LPC\lowercase{s}  discovered over the 1801--2017 period.}\label{sec:stat}

We use the JPL Small Body Database Search Engine\footnote{publicly available at https://ssd.jpl.nasa.gov/\_query.cgi} to construct a complete list of LPCs  discovered since 1801, omitting sungrazing comets  (the limit of 0.07\,au for the smallest accepted perihelion distance was adopted). Currently (August 2018), about 1500 orbits of sungrazing comets can be found in  JPL database, from among over 3\,000 already discovered \citep{Jones_et_al:2018}. Due to extremely short data-arcs of sungrazers (typically 1--3 days) almost all their orbits are obtained with the assumption of $e=1$; about $\sim$1480\,parabolic orbits with $q<0.07$\,au could be found in JPL database in August 2018. Most of the sungrazers were discovered by the Solar and Heliospheric Observatory (SOHO) over the period of 1995--2008. Sungrazers typically have only dozens of meters in diameter and evaporate during the observed perihelion passage. More than 80~per cent of them are members of the Kreutz group and are believed to be fragments of a one large comet that broke up several centuries ago \citep{Kreutz:1888,Sek-Chodas:2004, Sek-Chodas:2007}. A few other groups of sungrazers were also recognized, more numerous are the Kracht, Marsden and Meyer groups. Thus, this family of objects (often dynamically related to each other) requires a completely different approach than the parabolic comets with larger perihelion distances analysed in this paper, and therefore are not considered here.

The JPL database is the best source of data for such statistical analysis due to its completeness. However, it offers only osculating cometary orbits in a heliocentric frame. Basing on these orbits the three following lists of LPCs are available in JPL database:
\begin{itemize}
\item 'parabolic' comets,
\item 'hyperbolic' comets (in means of the shape of heliocentric osculating orbit),
\item 'other' comets; it includes comets with highly elliptical osculating orbits which are not matching any of the remaining  cometary orbit classes listed in the JPL database. 
\end{itemize}

 This JPL division into the last two subclasses of LPCs ('hyperbolic' and 'other' comets) is of little use from the point of view of this research. Instead, we will divide all these LPCs with known eccentricities into two other subgroups depending on the length of their original semimajor axis (section~\ref{sec:stat_LPCs}).

	\begin{table*}
		\caption{Eccentricities (column 6) and $1/a$-original (column 7) for some parabolic comets in JPL, where $1/a_{\rm ori}$ is given in units of $10^{-8}$\,au$^{-1}$
		\newline Column 8: 'Oortspike' denotes comets with $a_{\rm ori}>10^4$\,au, 'outside' means nominal orbit with $a_{\rm ori}<10^4$\,au within 1$\sigma$error, 'outside-' means that nominal orbit suggests $a_{\rm ori}<10^4$\,au, however within 1$\sigma$error this comet can also be an Oort spike comet, and 'HYP' marks hyperbolic original nominal orbit, however, additional '-' emphasizes that within  3$\sigma$error this comet can be inside Oort spike. 
	\newline KAT1901--1950 means PartII of Catalogue one-apparition comets (Kr{\'o}likowska et al, in prep.) 
	\newline Solution for C/1968~Q1, column 4: an asterisk notes that the last observation was taken with the weight of 0.1.
	\newline Second solution for C/2014~W$_{10}$, last column: $^{**}$ -- see text for details.} \label{table:parabolic}
		\centering
\setlength{\tabcolsep}{1.8pt} 	
		\begin{tabular}{cccccccccc}
			\hline
Name         & $q$   & arc  & No of obs.& RMS     &  e  & $1/a_{\rm ori}$ & nominal & orbital & References/ \\
             & [au]  &[days]& No of res.& [arcsec]&     &                 & orbit   & class   & Remarks     \\
$[$1$]$    & $[$2$]$ & $[$3$]$ & $[$4$]$& $[$5$]$ & $[$6$]$ & $[$7$]$     & $[$8$]$ & $[$9$]$ & $[$10$]$    \\
 \hline 
 &&&&&&&\\
 \multicolumn{10}{c}{{\bf Part (a)~~~} Years 1901--1950, parabolic comets in JPL, only comets with positive $1/a_{\rm ori}$ are shown} \\
C/1917 H1 & 0.764 &  68 &   188/319 & 3.47 & 0.997884$\pm$0.000880 &  1629$\pm$1203   & outside & 3a & KAT1901--1950 \\       
C/1923 T1 & 0.778 &  55 &    30/57  & 2.80 & 1.000149$\pm$0.000132 &   273$\pm$170    & outside & 3a & KAT1901--1950 \\
C/1925 X1 & 0.323 &  41 &    54/94  & 1.63 & 1.000048$\pm$0.000106 &     3$\pm$334    & Oortspike& 3a & KAT1901--1950 \\
C/1927 A1 & 1.036 &  76 &    39/74  & 2.69 & 1.000212$\pm$0.000357 &    93$\pm$342    & Oortspike& 2b & KAT1901--1950 \\
C/1930 L1 & 1.152 &  50 &    61/103 & 4.44 & 0.996257$\pm$0.002128 &  3442$\pm$1828   & outside & 3a & KAT1901--1950 \\
C/1932 H1 & 2.331 &  86 &    94/164 & 2.33 & 1.002012$\pm$0.000500 &   214$\pm$214    & outside-& 2b & KAT1901--1950 \\
C/1935 M1 & 3.502 &  54 &    19/33  & 1.83 & 0.988430$\pm$0.010129 &  3947$\pm$2865   & outside & 3a & KAT1901--1950 \\
C/1946 K1 & 1.018 &  50 &   113/180 & 4.92 & 0.996570$\pm$0.002713 &  4175$\pm$2643   & outside & 3a & KAT1901--1950 \\
C/1947 F2 & 0.962 &  43 &    52/86  & 2.92 & 0.998737$\pm$0.001612 &  1389$\pm$1681   & outside-& 3a & KAT1901--1950 \\
C/1947 K1 & 1.403 &  80 &    28/46  & 2.87 & 0.998553$\pm$0.001836 &  1512$\pm$1305   & outside & 3a & KAT1901--1950 \\
&&&&&&&\\
 \multicolumn{10}{c}{{\bf Part (b)~~~}Year 1951-2017, all parabolic comets in JPL with data-arc longer than 4 weeks} \\
\rowcolor{LightGray1}C/1951 C1 a1 & 0.719 &  92 &    19/33  & 1.02 & 1.000620$\pm$0.000085 &    -240$\pm$119 & HYP-     & 2a & present study\\
C/1962 H1 & 0.653 &  63 &    14/25  & 0.72 & 0.999762$\pm$0.000096 &       646$\pm$147  & outside  & 2b & present study\\
C/1965 S2 & 1.294 &  30 &    29/52  & 0.37 & 0.998269$\pm$0.000984 &      1167$\pm$762  & outside  & 3a & present study\\
C/1966 R1 & 0.881 &  30 &    25/47  & 1.32 & 0.998594$\pm$0.002852 &      2657$\pm$3234 & outside- & 3b & present study\\
C/1967 C2 & 1.327 &  59 &    55/104 & 1.68 & 0.996853$\pm$0.001420 &      2958$\pm$1053 & outside  & 3a & present study\\
\rowcolor{LightGray1}C/1967 M1 & 0.178 &  66 &    43/74  & 0.96 & 1.000239$\pm$0.000076 &    -977$\pm$428 & HYP-     & 2b & present study\\
C/1968 Q1 & 1.771 & 226 &    54$^*$/107 & 1.51 & 0.998610$\pm$0.000167 &    1144$\pm$94 & outside  & 2a & present study\\
\rowcolor{LightGray1}C/1968 Q2 & 1.099 &  75 &    35/67  & 1.06 & 1.001209$\pm$0.000221 &    -216 $\pm$201 & HYP-     & 2b & present study\\
C/1968 U1 & 2.599 &  31 &    18/36  & 1.42 & 0.993456$\pm$0.004634 &      2641$\pm$1785 & outside  & 3a & present study\\
\rowcolor{LightGray1}C/1969 P1 & 0.774 &  75 &    75/142 & 1.76 & 1.000220$\pm$0.000146 &       -10$\pm$192& HYP-     & 2b & present study\\      
\rowcolor{LightGray1}C/1970 U1 & 0.406 &  38 &    51/101 & 1.77 & 1.000620$\pm$0.000592 &      -706$\pm$1442 & HYP-   & 3a & present study\\
C/1977 H1 & 1.118 &  30 &    43/83  & 2.56 & 0.999978$\pm$0.001704 &       151$\pm$1535 & outside- & 3a & present study\\
C/1978 C1 & 0.437 &  35 &    75/147 & 1.21 & 0.999550$\pm$0.000098 &      1262$\pm$222  & outside  & 3a & present study\\
C/1978 T3 & 0.429 &  41 &    16/31  & 2.44 & 0.989142$\pm$0.001268 &     25572$\pm$3012 & outside  & 3b & present study\\
\rowcolor{LightGray1}C/1980 O1 & 0.525 &  42 &    25/45  & 2.39 & 1.003894$\pm$0.002006 &     -7223$\pm$3802 & HYP-    & 3b & present study\\
\rowcolor{LightGray1}C/1985 K1 & 0.106 & 102 &    40/71  & 1.95 & 1.000029$\pm$0.000051 &      -188$\pm$481  & HYP-    & 2b & present study\\
C/1987 T1 a1 & 0.514 &  38 &    25/49  & 3.40 & 0.996937$\pm$0.002020 &      6639$\pm$3926 & outside  & 3b & \\
\rowcolor{LightGray1}C/1987 W1 & 0.199 &  31 &    54/101 & 2.04 & 1.003913$\pm$0.000819 &    -19600$\pm$4193 & HYP     & 3b & present study\\
\rowcolor{LightGray1}C/1988 C1 & 1.931 &  56 &    23/40  & 0.56 & 1.002555$\pm$0.000933 &      -585$\pm$482 & HYP-     & 2b & present study\\
C/1988 Y1 & 0.428 &  56 &    60/112 & 1.81 & 0.999624$\pm$0.000340 &      1408$\pm$779  & outside  & 3b & \\
\rowcolor{LightGray1}C/1989 A6 & 2.206 &  59 &    28/55  & 1.05 & 1.002107$\pm$0.000474 &      -260$\pm$216 & HYP-     & 2b & present study\\
C/1989 R1 & 1.324 &  55 &    19/37  & 1.80 & 0.987644$\pm$0.006636 &      9479$\pm$5094 & outside  & 3a & present study\\
C/1989 Y2 & 1.975 &  51 &    53/102 & 2.47 & 0.995662$\pm$0.002659 &      2635$\pm$1339 & outside  & 3a & present study\\
C/1990 E1 & 1.068 &  73 &    93/179 & 1.36 & 0.998476$\pm$0.000191 &      2120$\pm$180  & outside  & 2b & present study\\ 
C/1992 J2 & 0.592 &  31 &    16/30  & 2.60 & 0.999541$\pm$0.002833 &      1249$\pm$4761 & outside- & 3b & present study\\
C/2005 J2 & 4.289 & 357 &    68/128 & 0.69 & 1.000756$\pm$0.000066 &     458.8$\pm$15.4 & outside  & 1b & also in MPC\\
C/2005 L2 & 3.194 & 292 &    54/107 & 0.70 & 0.999742$\pm$0.000020 &     439.3$\pm$6.4  & outside  & 1b & also in MPC\\
C/2006 X1 & 6.126 &  86 &    96/185 & 0.39 & 0.999399$\pm$0.000251 &     185.1$\pm$41.3 & outside  & 2a & KD2017 \\
C/2007 Q1 & 3.010 &  24 &    43/84  & 0.58 & 1.003060$\pm$0.002232 &        55$\pm$799  & Oortspike& 3a & KD2013 \\ 
C/2009 G1 & 1.129 & 113 &   128/245 & 0.58 & 0.997625$\pm$0.000040 &      2477$\pm$36   & outside  & 2a & also in MPC\\
C/2014 C2 & 0.512 &  65 &   166/297 & 0.61 & 0.999675$\pm$0.000037 &      1398$\pm$71   & outside  & 2b & also in MPC\\
\rowcolor{LightGray1}C/2014 W$_{10}$ & 7.631 & 55 & 25/45 & 0.20 & 1.003760$\pm$0.001917 &  -493$\pm$252 & HYP-  & 3a & present study, upgraded in the revised version\\
\rowcolor{LightGray1}C/2014 W$_{10}$ & 7.582 & 28 & 17/34& 0.27 & 1.370485$\pm$0.379311 &    -48464$\pm$54882& HYP-     & 3b & present study, shorter arc$^{**}$\\
		\end{tabular}
	\end{table*}

\subsection{Parabolic comets}\label{sec:stat_par}

\begin{figure}
\includegraphics[width=8.9cm]{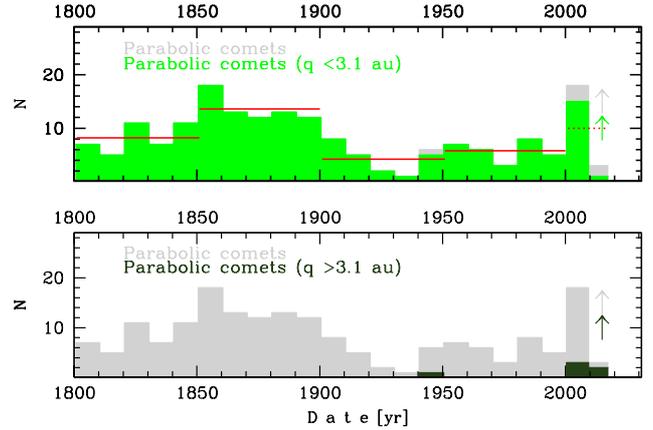} 
\protect\caption{\label{fig:stat_par}  Distribution of number of parabolic comets in the function of the moment of their discovery. All bins covered 10-yr period except the last bin starting from 2011 which spans a time period of 7\,yrs. In the upper panel the distribution of small-perihelion comets ($q<3.1$\,au, green histogram) is given on the gray background distribution of all parabolic comets, darkgreen histogram in the lower panel shows a small subsample of large-perihelion comets ($q>3.1$\,au).  The red horizontal lines in the upper panel show the mean rate of 'parabolic' comets discovery (mean number per decade) averaged over the 50-year periods. Vertical arrows drawn above the last bin (years 2011--2017) indicate that statistics will be significantly richer after 2010} 
\end{figure}

In June 2018, the original JPL list of parabolic comets consisted of 217~objects having perihelion distances larger than 0.07\,au,  however, we have removed four secondary components (C/1860~D1-B, C/1994~G1-B, C/2005~K2-A, C/2011~J2-C), and C/2002~Q3 because its main component A is in the list of 'other' comets. Additionally, we have moved ten parabolic comets discovered during the 1901-1950 period (C/1917~H1, C/1923~T1, C/1925~X1, C/1927~A1, C/1930~L1, C/1932~H1, C/1935~M1, C/1946~K1, C/1947~F2, C/1947~K1) to the remaining two cometary lists according to orbital results obtained by Kr{\'o}likowska et al. (in prep., see the part~(a) of Table~\ref{table:parabolic}). 

Additionally, we have tried to determine definitive orbits for some other 'parabolic' comets. Unfortunately, as many as 109 comets from this list are objects discovered before 1900 and there are not yet available in contemporary databases. However, such data are available at the Minor Planet Center for the majority of comets discovered after 1950. That is why we have focused only on this group of comets and selected objects observed longer than 4~weeks. The choice of this particular criterion is obviously a simplification, because in addition to the length of the observation interval, many other factors affect the quality of an orbit: number of observations and the accuracy of a single measurement, location of the data-arc along the orbit (how far they are from a perihelion), and a distance of a perihelion from the Sun.

As a result, we found 30 comets (37 per cent of all 'parabolic' comets discovered since 1950 and presented in the JPL database in July 2018) and calculated their definitive orbits. Characteristics of the observational material and obtained eccentricity of osculating orbit as well as $1/a$ for the original orbit are given in part~(b) of Table~\ref{table:parabolic} (columns 1--8), together with additional two comets basing on our previous results -- see below). Column~9 gives a quality of the obtained orbits according to the method proposed by us in KD2013.  

It is worth to mention an interesting case of the most hyperbolic original orbit of C/1987~W1. In the first version of this paper we obtained even more hyperbolic original orbit for another comet, C/2014~W$_{10}$ (see the second row for this comet in Table ~\ref{table:parabolic}). In the original version of this manuscript a relatively high eccentricity of $1.37\pm 0.38$ was given by us. As a result of such an unusual orbital solution a call for a search for additional data was announced \citep{dyb-kroli_w10:2018}. After this astro-ph note publication a few more and corrected data were obtained by us via a private communication with Robert Weryk\footnote{This observations will be published in the Minor Planet Circular, to be issued in December 2018} and we were able to upgrade this unusual solution (see first orbit for this comet in Table~\ref{table:parabolic}). There are still some unresolved problems with the orbit of this comet and we plan to update the above mentioned note in the future. 

It should be stressed, that the Oort spike defined strictly as $0 < 1/a_{\rm ori} < 100$ in units of $10^{-6}$\,au$^{-1}$ is very narrow comparing with typical $1/a_{\rm ori}$ uncertainties presented in Table~\ref{table:parabolic}. As a result only one comet, C/2007~Q1, can be formally classified as the Oort spike member in a strict sense, 11~comets as originally hyperbolic and the rest as highly elliptical.

Instead of that formal description of Oort spike we propose a softened one which takes into account large $1/a_{\rm ori}$ uncertainties and indeterminacy of the NGF in these cases. Column~8 describes this classification of the semimajor axis of original orbit. We introduced here three types of original orbits:
\begin{enumerate}
\item When nominal orbits have $a<10\,000$\,au we remark them as 'outside' the Oort spike if the value of $1/a_{\rm ori}$ is farther than $1\sigma$-error from the Oort spike right border ($1/a_{\rm ori}=100$ in units of $10^{-6}$\,au$^{-1}$) and as 'outside-' if this nominal value is closer than $1\sigma$-error. We have found 17~objects belonging to the first group and 3~comets to the second group.
\item When nominal orbits have $a>10\,000$\,au we remark them as 'Oort spike'. We found only one such object.
\item All comets highlighted in grey are comets with hyperbolic original nominal orbit. These orbits within 3$\sigma$-uncertainties are compatible with Oort spike orbits, except the orbit of C/1987~W1. However, we can only speculate that all 'HYP' orbits presented in Table~\ref{table:parabolic} can be counted as orbits of comets coming from the outermost part of the Oort Cloud, therefore we retain them in the list of parabolic comets. We have obtained 11~such objects.
\end{enumerate}

We obtained that only 40~per cent of analysed sample of 30 'parabolic' comets discovered since 1950 seems to have $a_{\rm ori}>10\,000$\,au.  Basing on the above results, we can conclude that the whole sample of 'parabolic' comets probably contains Oort spike comets (in a wider sense, i.e. also slightly hyperbolic), and LPCs with more tight orbits in more or less equal proportions.

According to the above results ((part~(b) of Table~\ref{table:parabolic}) we moved 15~comets (C/1962~H1, C/1965~S2, C/1966~R1, C/1967~C2, C/1968~Q1, C/1968~U1, C/1977~H1, C/1978~C1, C/1978~T3, C/1987~T1, C/1988~Y1, C/1989~R1, C/1989~Y2, C/1990~E1, C/1992~J2) to the 'other' list of comets.  Two more, C/2006~X1 and C/2007~Q1, were shifted to the 'other' and 'hyperbolic' lists, respectively on the basis of our previous results given in KD17 and KD13, respectively. We found that other four comets (C/2005~J2, C/2005~L2, C/2009~G1, C/2014~C2) have definitive  orbits in MPC (see last column in Table~\ref{table:parabolic}), and according to our and MPC results the first was included in a 'hyperbolic' list and the remaining three in 'other' comets.  This way, the original JPL list containing 212 different 'parabolic' comets has been reduced here by 15 per cent. 

\begin{figure}
\includegraphics[width=8.9cm]{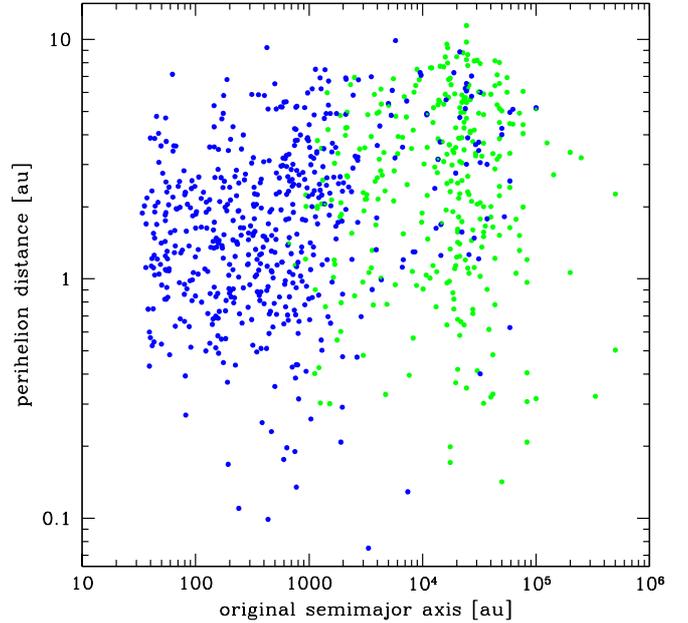} 
\protect\caption{\label{fig:JPL_two_lists}  Perihelion distances and original semimajor axes for 842 LPCs with definitive orbits ($e\ne 1$, LPCs with hyperbolic original orbits are excluded). Comets having hyperbolic osculating orbits ('hyperbolic' list of comets in JPL database) are shown with green dots while LPCs with elliptical osculating orbits ('comet (other)' list of LPCs in JPL) are marked with blue dots.} 
\end{figure}

\begin{figure*}
\includegraphics[width=7.8cm]{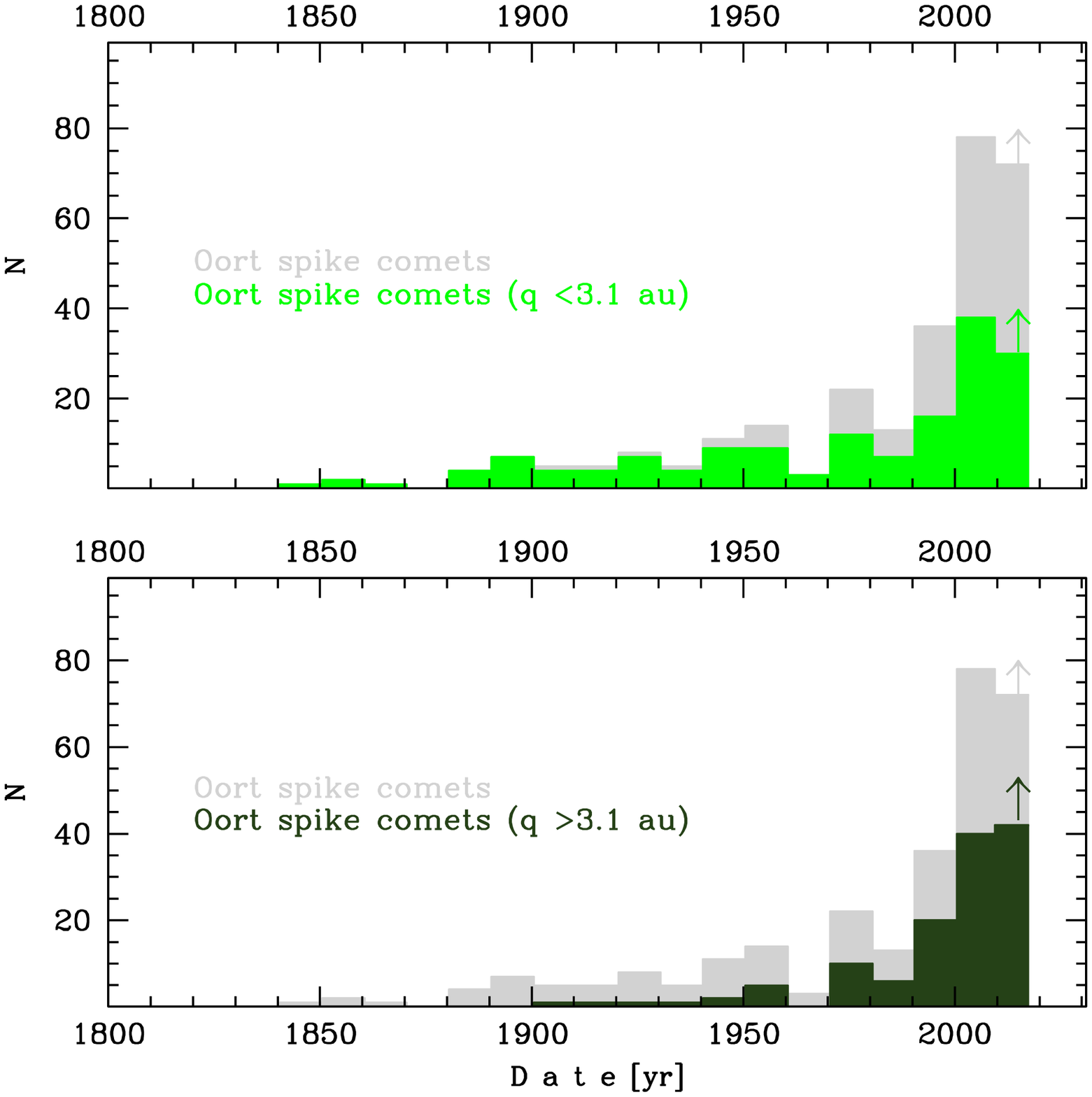} 
\includegraphics[width=7.8cm]{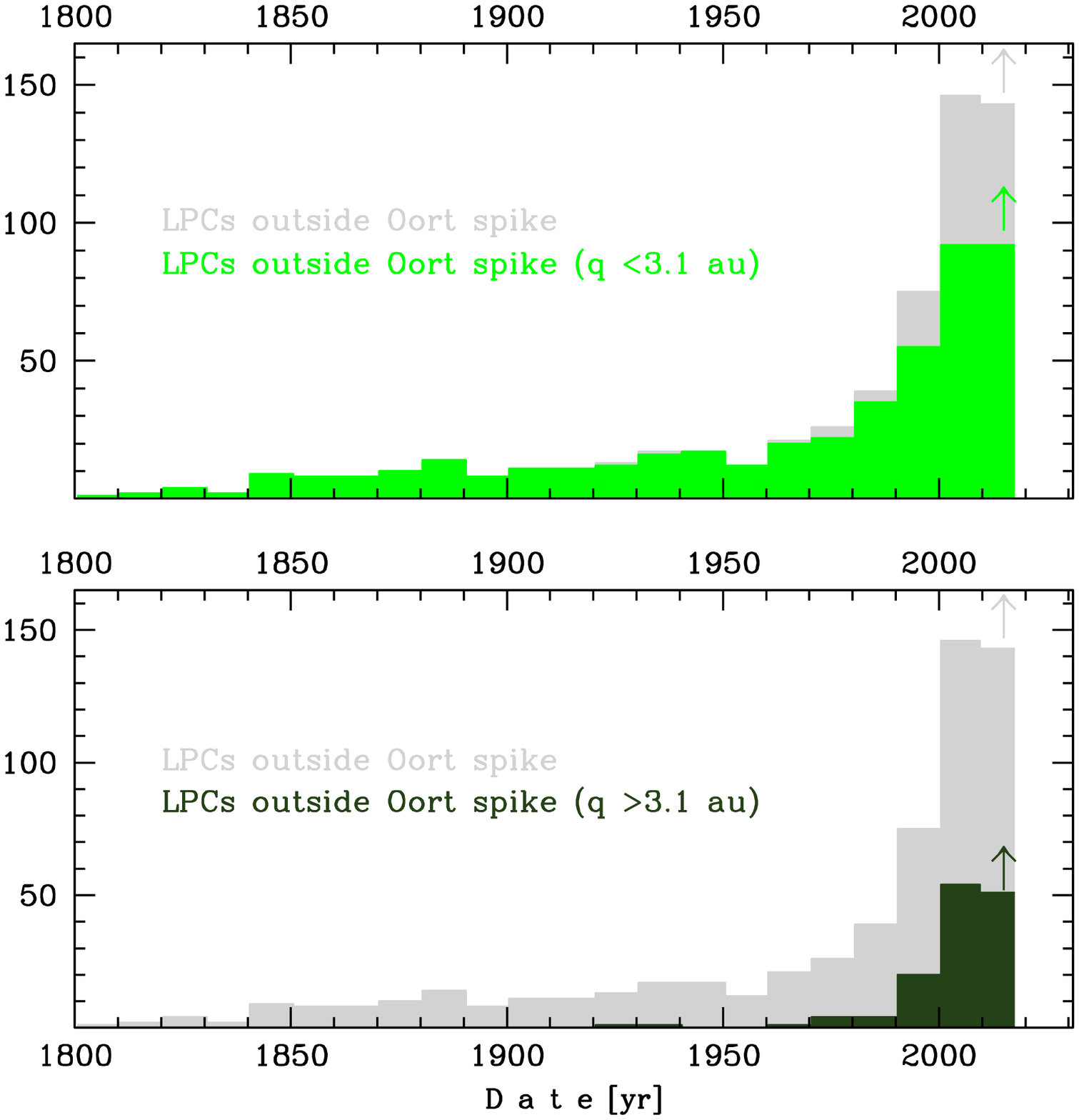} 
\protect\caption{\label{fig:Oort_and_other}  Distribution of Oort spike comets (left-side panels) and the LPCs with more tight orbits (right-side panels) in a function of the moment of their discovery. See text for a detailed explanation.} 
\end{figure*}

Finally, a sample of 181 'parabolic' comets was  taken into account. Distribution of their discovery throughout the analysed period is given in Fig.~\ref{fig:stat_par}. Only six of them have a perihelion distance larger than  3.1\,au (dark-green histogram in the bottom panel). Four horizontal red lines show the mean rate of discoveries averaged over the 50-yr intervals, and the fifth dotted red line -- averaged over the last 17\,yr time-interval. During the period of 1901--1950 two world wars took place which might have caused a decrease in comet discoveries during that period. This caused a wavy trend with deep minimum  between 1901--1950 (third red horizontal line from the left). The general trend in the significantly smaller number of parabolic comets over the period of 1950--2000 (mean discovery rate of 5.8~per decade) compared to the 1851--1900 period  (13.6 comets per decade) most likely results from improved observation capabilities and incomparably better numerical calculations capabilities (computers) in the second half of 20th century. 
It seems that the high discovery rate of parabolic comets for the 1851--1900 period is seriously overestimated here compared to more recent periods. Once again it is worth noting that the list of parabolic comets has been shortened by us only since 1901. We have checked that in the period 1851--1900, 8~'parabolic' comets were observed longer than three months, 22~objects longer than two months, and as many as 41~comets longer than one month. Let's assume that for 50~per cent of these 41~comets it would be possible to determine a definitive orbit with positive original semimajor axis (just as for the 1951--2000 period, see part(b) of Table~\ref{table:parabolic}). Then this rate would be only about 9.2~of 'parabolic' comets per decade within the second half of 19th century. Unfortunately, to do this it would be necessary to collect the positional data from the huge number of original papers and then process (comet-star)-measurements through modern star catalogues \citep[see for example][]{kroli-dyb:2016} for the majority of these comets.This work is in progress.

\begin{figure}
\includegraphics[width=8.9cm]{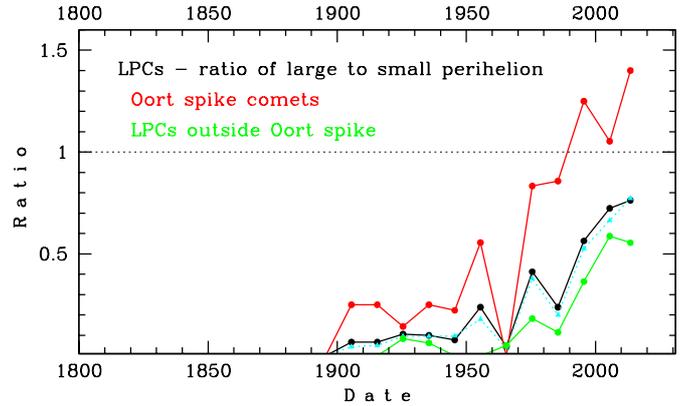} 
\protect\caption{\label{fig:ratio}  Number ratio of large-perihelion comets ($q>3.1$\,au) to small-perihelion ones ($q<3.1$\,au) in function of a discovery period. Red curve shows statistics for Oort spike comets whereas the green curve describes LPCs outside Oort spike, the overall ratio for the whole sample of LPCs is plotted in black (parabolic comets were not included in that sample) and cyan colour (parabolic comets were included). These ratios were calculated for 10-yr bins, except for the last bin which covers 7\,yr-period from 2011 to 2017.} 
\end{figure}

\noindent On the other hand, in the first decade of 21th century as much as 14 'parabolic' comets were observed shorter than two weeks, and next four shorter than four weeks. The majority of them were discovered automatically as very dim objects and for that reason sometimes classified at first as asteroids. Moreover, almost all of them were discovered just within  a few days around their perihelion.

Generally, parabolic comets constitute a large part of LPCs (around 17~per cent of all considered LPCs after all our reclassifications). As was mentioned above, we determined definitive orbits for 32~previously parabolic comets in JPL database including four comets which have also definitive orbits in IAU Minor Planet Center database presented in the part~(b) of Table~\ref{table:parabolic}. We also added 10~objects with definitive orbits taken from Kr{\'o}likowska et al. (in prep., part~(a) of Table~\ref{table:parabolic}). Statistics are not impressive, however in this subsample of 42 'improved' orbits there are 3~Oort spike comets, 28~LPCs moving on more tight orbits ($a_{\rm ori}<10\,000$\,au), and 11~comets probably from the Oort spike in a wider sense. It is reasonable to assume that it is more or less similar within the entire sample of parabolic comets, where definitive orbits can not be determined. Thus, parabolic comets (97~per cent of them have small perihelion distances and consequently probable significant NGF) do not seem to be able to change the general conclusions that will be drawn further in this paper from the comparative analysis of Oort spike comets and remaining LPCs with definitive orbits. Under that last term, we mean all LPCs with determined eccentricities from the positional data, in short all LPCs with $e\ne 1$.

\begin{table*}
\caption{Statistic of various types of LPCs discovered in different periods. This statistic has been prepared using the following sources: KD13, KD17, \protect\cite{krolikowska:2014,krol-sit-et-al:2014}, JPL Small-Body Database Search Engine (May~2018) and Minor Planet Center Database Search (May 2018). \textit{Parabolic comet} means a comet having an orbit determined with the assumption that $e=1$ (comets with very poorly known orbits). \textit{Oort spike comets} are comets having
$a_{\textnormal{ori}} > 10\,000$\,au whereas \textit{remaining comets} include all LPCs with
$a_{\textnormal{ori}} < 10\,000$\,au.}\label{table:statistics} 
		\centering
		\begin{tabular}{ccccccccc}
\hline
			& \multicolumn{2}{c}{Parabolic comets}  & \multicolumn{2}{c}{Oort spike comets} & \multicolumn{2}{c}{Remaining LPCs} &    \multicolumn{2}{c}{All three types together}   \\
			years & $q<3.1$au    & $q>3.1$au  & $q<3.1$au    & $q>3.1$au  & $q<3.1$au    & $q>3.1$au  & $q<3.1$au    & $q>3.1$au  \\
\hline 
			&&&&&&&& \\
			1801--1810 &  7      & 0          &  0           &  0         &   1          &  0         &  8           &   0        \\
			1811--1820 &  5      & 0          &  0           &  0         &   2          &  0         &  7           &   0        \\
			1821--1830 & 11      & 0          &  0           &  0         &   4          &  0         & 15           &   0        \\
			1831--1840 &  7      & 0          &  0           &  0         &   2          &  0         &  9           &   0        \\
			1841--1850 & 11      & 0          &  1           &  0         &   9          &  0         & 21           &   0        \\
			1851--1860 & 18      & 0          &  2           &  0         &   8          &  0         & 28           &   0        \\
			1861--1870 & 13      & 0          &  1           &  0         &   8          &  0         & 22           &   0        \\
			1871--1880 & 12      & 0          &  0           &  0         &  10          &  0         & 22           &   0        \\
			1881--1890 & 13      & 0          &  4           &  0         &  14          &  0         & 31           &   0        \\
			1891--1900 & 12      & 0          &  7           &  0         &   8          &  0         & 27           &   0        \\
			&&&&&&&& \\
			1901--1910 &  8      & 0          &   4          &  1         &   11         &  0         & 23           &   1        \\
			1911--1920 &  5      & 0          &  4           &  1         &  11          &  0         & 20           &   1        \\
			1921--1930 &  2      & 0          &  7           &  1         &  12          &  1         & 21           &   2        \\
			1931--1940 &  1      & 0          &  4           &  1         &  16          &  1         & 21           &   2        \\
			1941--1950 &  5      & 1          &   9          &  2         &   17         &  0         & 31           &   3        \\
			1951--1960 &  7      & 0          &  9           &  5         &  12          &  0         & 28           &   5        \\
			1961--1970 &  6      & 0          &  3           &  0         &  20          &  1         & 29           &   1        \\
			1971--1980 &  3      & 0          & 12           & 10         &   22         &  4         & 37           &  14        \\
			1981--1990 &  8      & 0          &  7           &  6         &  35          &  4         & 50           &  10        \\
			1991--2000 &  5      & 0          & 16           & 20         &  55          & 20         & 76           &  40        \\
			&&&&&&&& \\
			2001--2005 &  9      & 0          & 16           & 22         &  41          & 25         & 67           &  47        \\
			2006--2010 &  6      & 3          & 22           &  18        &  51          & 29         & 79           &  50        \\
			2011--2015 &  1      & 2          &  24          &  29        &   65         & 38         &  90          &  69        \\
			2016--2017 &  0      & 0          &   6          &  13        &   27         & 13         &  33          &  26        \\
			&&&&&&&& \\
			\rowcolor{yellow}2001--2017 & 16  & 5  & 68      &  82        &  184         & 105        & 269          & 192        \\
			\rowcolor{yellow}1901--2000 & 50  & 1  &  75     & 47         &   211        &  31        & 336          & 79         \\
			\rowcolor{yellow}1801--1900 & 109 & 0  & 15      &  0         &   66         &   0        & 190          & 0          \\
			&&&&&&&& \\
			1801--2017 & 175     & 6          &  158         &  129       &   461        &  136       & 795          & 271        \\
			& \multicolumn{2}{c}{181 }  & \multicolumn{2}{c}{ 287 } & \multicolumn{2}{c}{ 597 } &  \multicolumn{2}{c}{1065} \\
			\hline
		\end{tabular}
	\end{table*}

\subsection{Sample of LPCs with definitive orbits}\label{sec:stat_LPCs}

The original list within 'hyperbolic' class of orbits in JPL database consisted (in June~2018) of 337~comets discovered in the  1801--2017 period, where 5 of them are the secondary cometary components excluded by us. We added $1/a_{\rm ori}$-values to the data using our previous original orbit determinations for Oort spike comets defined by $1/a_{\rm ori}<0.000100$\,au$^{-1}$:
\begin{itemize}
\item from \cite{krol-sit-et-al:2014}: comets from the years between 1901--1950,
\item from KD17: comets with $q>3.1$\,au, years between 1951--2010, 
\item from KD13: comets with $q<3.1$\,au, years 2001--2010,
\item from Dybczy{\'n}ski and Kr{\'o}likowska (in prep.): comets with $q>3.1$\,au, years 2011--2017), and
\item from Kr\'olikowska et al. (in prep.): a few comets from the years 1901--1950 with parabolic orbits in MWC\,08.
\end{itemize}
For the remaining comets we took values of  $1/a_{\rm ori}$ available thanks to IAU Minor Planet Center database and Nakano Notes. It turned out that 236~objects are Oort spike comets (in a wider sense i.e. including 42 slightly hyperbolic original orbits) and 96~comets have more tight original orbits from these among JPL 'hyperbolic' group.

Original JPL list of 'other' comets consisted of 511 objects. However, we excluded 153P/Ikeya-Zhang as a unique comet with orbital period longer than 200\,yr which was observed in its two consecutive perihelion passages. 
Additionally, we excluded C/2017~Y2 PANSTARRS which turned out to be a short-period comet according to Minor Planet Center (short arc of 19 days and poorly known orbit). Three comets  which are secondary components were also excluded. We collected the $1/a_{\rm ori}$-values for the 506~remaining 'other' comets from the same sources as those listed for the case of 'hyperbolic' list.  

\noindent Next, we found 14~LPCs defined in the JPL database as Jupiter-family comets (JFCs) according to the Levison and Duncan criterion which is based on the Tisserand parameter relative to Jupiter in the restricted circular three-body problem, $T_{\rm Jupiter}$, see for example \citet{levison:1996}. They define NIC (nearly isotropic comets) using the condition $T_{\rm Jupiter}<2$. For example, C/2004~P1 NEAT having $2 < T_{\rm Jupiter} < 3$, and semimajor axis about 8100\,au was classified at JPL as JFC. Another example is C/2016~X1 with semimajor axis of about 3240\,au which according to conditions: $T_{\rm Jupiter} > 3, \>a > a_{\rm Jupiter}$ is included in JPL in the group of Chiron-type comets. Both these comets, and  other 13~comets (C/1996~R3, C/1998~M1, C/2000~Y2, C/2002~P1, C/2009~U5, C/2012~LP$_{26}$, C/2013 G8, C/2013~TW$_5$, C/2013~U2, C/2014~M2, C/2014~W8, C/2015~J2, C/2016~P4) having orbital period longer than 200\,yr are included as LPCs in the present analysis. This increased the sample to 521~objects. It turned out that 48 objects from this list are Oort spike comets and 473 are LPCs moving on more tight orbits. 

At the end, we combined both these lists, and added 31~comets which we removed from the original 'parabolic' list as comets with known eccentricities (see the previous section and Table~\ref{table:parabolic}).

Finally, our sample of LPCs discovered in the years 1801--2017, and with known eccentricities and perihelion distances larger than 0.07\,au, consists of 884 comets. 

Fig.~\ref{fig:JPL_two_lists} shows perihelion distances of all these LPCs with positive $1/a$-original values (barycentric frame) where comets having hyperbolic osculating heliocentric orbits are represented as green dots whereas the remaining LPCs are given by blue dots. It turned out that only 42 comets formally have hyperbolic barycentric original orbits and were excluded from this plot. However, in most cases orbits of these 42~comets can be elliptical within 3$\sigma$ uncertainties, see the discussion on this in the Introduction. 

Similar plot,  using also logarithmic scales is presented in \citet{Dones:2015}, but for osculating heliocentric semimajor axes (see Fig.~3 therein). As a result all 'hyperbolic comets' in term of their osculating orbits (green dots in our Fig.~\ref{fig:JPL_two_lists}) are omitted in their figure. Additionally, in the figure based on the osculating heliocentric semimajor axis it is impossible to clearly indicate where Oort spike comets are located. Moreover, this way of presentation resulted in the illusory impression that semimajor axes of LPCs exceed 10\,000\,au only for a very small number of LPCs. We can not find any information how the sample of 1188~comets given in Fig.~3 of \citet{Dones:2015} was selected. However, such a selection gives a poorly representative sample of known LPCs (it is important to note that all parabolic comets are  omitted in both plots).  

\begin{center}
	\begin{table}
		\caption{Ratio of the number of large-perihelion comets to small-perihelion comets for Oort spike comets (in a wider sense) and for the remaining LPCs ($a<10\,000$\,au); see text for details.}\label{table:ratio_large_small}
		\centering
\setlength{\tabcolsep}{1.8pt} 	
		\begin{tabular}{ccccc}
\hline
  Years & \multicolumn{2}{c}{Ratio}                        &   \multicolumn{2}{c}{Ratio (with a correction }                           \\
        &  \multicolumn{2}{c}{(without parabolic comets)} &   \multicolumn{2}{c}{on parabolic comets)} \\
        & Oortspike   & Remaining  &  Oortspike & Remaining  \\                     
        &   comets    &  LPCs      &  comets    & LPCs  \\                     
\hline 
			&&&& \\
			1971--1980 &  0.83 & 0.18 & 0.75 & 0.17    \\
			1981--1990 &  0.86 & 0.11 & 0.55 & 0.10    \\
			1991--2000 &  1.25 & 0.36 & 1.08 & 0.35    \\
			2001--2010 &  1.05 & 0.59 & 0.90 & 0.55    \\
			2011--2017 &  1.40 & 0.55 & 1.39 & 0.57    \\
			\hline
		\end{tabular}
	\end{table}
\end{center}

Distribution of LPCs discovery throughout the analysed period is given in Fig.~\ref{fig:Oort_and_other} and Table~\ref{table:statistics}. 

Light-gray histogram visible in the left-side panels of Fig.~\ref{fig:Oort_and_other} shows distribution of all Oort spike comets (including hyperbolas) in the 10-yr bins (287 objects) whereas light-gray histogram given in the right-side panels represents distribution for the full sample of the remaining LPCs with definitive orbits (597 objects with $e\ne 1$). 

\section{Small perihelion versus large perihelion LPCs}\label{sec:stat_large_small}

Green histograms in the upper part of Fig.~\ref{fig:Oort_and_other} show similar statistics as above for small-perihelion comets in the same bins (158 and 461 comets, respectively for Oort spike comets and LPCs with more tight orbits) whereas dark-green histograms in the lower part of this figure represent analogous  distributions for large-perihelion comets (129 and 136 objects, respectively). 

Generally, the rapid increase in the number of discovery of all subgroups of LPCs with $e\ne 1$ is evident. This increase began around 1970 and has continued until now, with the remarkable discovery rate increase just before the end of the 20th century. It is worth to notice that the number of discoveries of small-perihelion LPCs in the last 17\,yr (2001--2017) is comparable to the number of discoveries of the objects of the same type during the whole 20th century (see Table~\ref{table:statistics}). Conversely, the number of large-perihelion comets discovered in this short period is more than twice as large as during the 20th century. It means that in the last 17\,yr we have observed a remarkable increase in the rate of large-perihelion LPCs discovery, mainly due to the effectiveness of large sky surveys.

C/1906~E1 was the first discovered comet with $q>3.1$~au in the sample studied in this paper but discoveries of large-perihelion LPCs were very rare before 1970 and only 15 objects with $q>3.1$\,au have been detected up to that time while around two hundred of small-perihelion comets were known by that time. Since 1970 the number of large perihelion LPCs has been systematically increasing. Currently, they constitute 30~per cent of the whole sample of LPCs  with $e\ne 1$. 

We noticed a surprising fenomenon that in the sample of Oort spike comets this percentage of large-perihelion comets is even significantly higher and reaches 45~per cent while in the remaining sample of LPCs on more tight orbits it equals only 23~per cent. 
This substantial difference in the percentage of large-perihelion comets in both samples (the Oort spike and LPCs with more tight orbits) is also easily visible in Fig.~\ref{fig:Oort_and_other} and Table~\ref{table:statistics}. 

Fig.~\ref{fig:ratio} additionally shows how the discovery ratio of large-perihelion comets to small-perihelion comets has evolved since 1800 in these different groups of LPCs. 

\noindent Presented statistics allow reliable conclusions only  for the period since 1970s (see  Table~\ref{table:statistics}). Fig.~\ref{fig:ratio} shows that since 1970 the ratio of the number of large-perihelion comets to the number of small-perihelion comets  has been higher for comets from the Oort spike (in a wider sense, i.e. all comets having $1/a_{\rm ori}< 0.0001$~au$^{-1}$, red curve) compared to the remaining LPCs ($a_{\rm ori}<10\,000$\,au, green curve). The black line shows this ratio for all LPCs considered here. 

The question is whether this effect would be visible if it  was possible to incorporate parabolic comets into both subsamples. 
Using the previous conclusions from the section~\ref{sec:stat_par} let us assume that parabolic comets would give a similar contribution to both groups of LPCs.
The result of such a simple assumption on how this ratio would change after 1970 is presented in Table~\ref{table:ratio_large_small}. It can be seen that these quite different proportions of large-perihelion comets to small-perihelion comets  still exist between both subsamples of LPCs for such a corrected ratio in all five time periods (compare column 4 and 5 in this table).

\section{Discovery distances}\label{sec:distances}

\begin{figure}
	\includegraphics[angle=270, width=9cm]{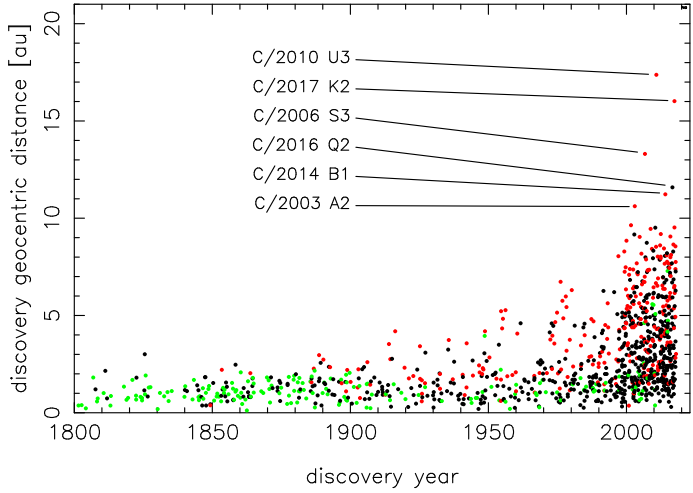} 
	\protect\caption{\label{fig:discovery_distances} Geocentric discovery distances for all 1065~LPCs versus discovery date. Green dots represent parabolic comets, red dots Oort spike comets in a wider sense ($1/a_{\rm ori}<0.0001$\,au$^{-1}$) and black dots depict the remaining LPCs.}   
\end{figure}

\begin{figure}
	\includegraphics[angle=270, width=9cm]{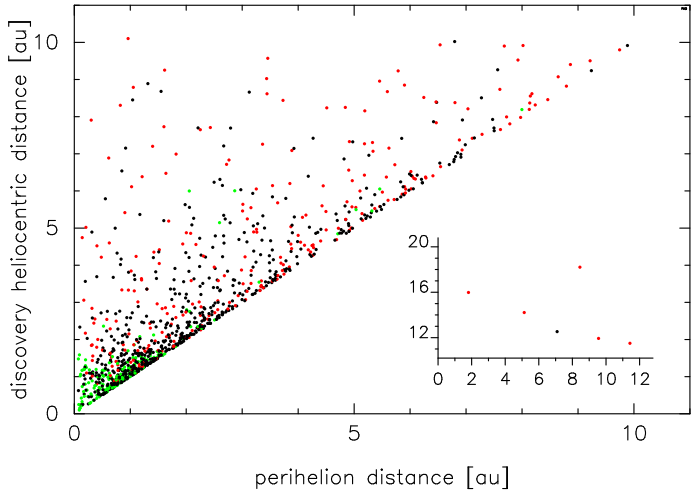} 
	\protect\caption{\label{fig:dist-versus-q} Heliocentric discovery distances versus the perihelion distances for all 1065~LPCs. Dot colour coding is as in Fig.~\ref{fig:discovery_distances}. Six comets discovered at exceptionally large distance are omitted in the main plot but presented in the inset at the right lower corner. These six comets are named in Fig.~\ref{fig:discovery_distances}. }   
\end{figure}

\begin{table}
\caption{Number of the Oort spike, parabolic and remaining LPCs discovered after 1970 in four discovery geocentric distance $\Delta_{\rm disc}$ intervals. Three separate statistics are shown: for all discovery elongations, for elongations greater than $90\degr$ and smaller than this threshold.  In the last row of each section of the table the respective median discovery distance is also given. }\label{table:discovery_distances}
\centering
\setlength{\tabcolsep}{1.8pt} 	
\begin{tabular}{ccccc}
	\hline
	$\Delta_{\rm disc}$  & Oort spike & Parabolic & Remaining & Number    \\
	interval [au]        & comets     & comets    & LPCs      & ratio     \\
	                     & $[$O$]$    & $[$P$]$   & $[$L$]$   & $[$O/L$]$ \\
	\hline 	
	&&&&  \\
	\multicolumn{5}{c}{All ~~~discovery ~~~elongations} \\
	&&&& \\
	0 -- 2   & 21  & 29 &  181 & 0.12 \\
	2 -- 5   & 88  &  6 &  195 & 0.45 \\
	5 -- 10  & 94  &  4 &   66 & 1.42 \\
   10 -- 20  &  5  &  0 &    1 & 5.   \\
	&&&&  \\
	median  $\Delta_{\rm disc}$ [au]: & 4.77 & 1.32 & 2.44 & \\
    &&&&  \\	
	\hline
	&&&&  \\
	\multicolumn{5}{c}{Elongations ~~~greater ~~~than ~~~90\degr} \\
	&&&&  \\
	0 -- 2   &  9  &  6 &   96 & 0.094  \\
	2 -- 5   & 68  &  6 &  151 & 0.45  \\
	5 -- 10  & 85  &  4 &   56 & 1.51  \\
	10 -- 20  &  5  &  0 &    1 & 5.     \\
	&&&& \\
	median  $\Delta_{\rm disc}$ [au]: & 5.35 & 2.33 & 2.95 \\
	&&&&  \\
	\hline
	&&&&  \\	
	\multicolumn{5}{c}{Elongations ~~~smaller ~~~than ~~~90\degr} \\
	&&&& \\
	0 -- 2   & 12 & 23 & 85 & 0.14 \\
	2 -- 5   & 20 &  0 & 44 & 0.45 \\
	5 -- 10  &  9 &  0 & 10 & 0.9  \\
	10 -- 20  & 0 &  0 &  0 & --   \\
	&&&& \\
	median  $\Delta_{\rm disc}$ [au]: & 3.19 & 1.17 & 1.73 &          \\
	&&&&  \\
	\hline
\end{tabular}
\end{table}

Searching for the interpretation of the phenomenon described above we decided to investigate  discovery circumstances, especially the discovery distances. With a great help of Meyer Catalogue of Comet
Discoveries (personal communication\footnote{This catalogue is available upon request send to maik@comethunter.de}) we collected all discovery dates for the considered LPCs and calculated geocentric and heliocentric distances and elongations for these epochs. In Fig.~\ref{fig:discovery_distances} we present geocentric discovery distances of all considered comets, divided into parabolic, Oort spike and the remaining LPCs. 
It is a striking fact visible in Fig.~\ref{fig:dist-versus-q} that after 2000 we discover comets at large distances independently of their perihelion distances.  
Figs.~\ref{fig:discovery_distances}-\ref{fig:dist-versus-q} show that the Oort spike comets (red dots) are discovered statistically at larger distances and more often have larger perihelion distances than LPCs on more tight orbits (black dots). We note that among LPCs discovered further than 7\,au as much as 72~per~cent are Oort spike comets. This percentage grows to 79 if we move to 8\,au.  Numbers describing the whole statistics after 1970 can be found in Table~\ref{table:discovery_distances}. 
The last column in this table additionally shows how the number ratio of Oort spike comets and remaining LPCs increases in a function of the geocentric distance at the moment of discovery, $\Delta_{\rm disc}$. For $\Delta_{\rm disc}$<2.0\,au this ratio is  smaller than 0.3 (under the conservative assumption for parabolic comets included in the group of Oort spike comets) whereas for $\Delta_{\rm disc}$>5.0\,au is about 1.5.

The actual distribution of the perihelion distance is still unknown, we can therefore only speculate about the most probable explanations for this observational fact, i.e. more numerous discoveries of Oort spike comets with large perihelion distance than other LPCs with large $q$. Generally, this may be the effect of significant differences in actual (unknown) perihelion distance distributions between Oort spike comets and LPCs outside the Oort spike (especially for large-perihelion comets, see also sect.~\ref{sec:perihelion}) or  it can be caused by a simple observational selection effect, reflecting  the fact, that  Oort spike comets are, for some reason statistically discovered at larger distances (both geocentric and heliocentric). Assuming that unknown $q$-distributions (for Oort spike comets and LPCs moving on more tight orbits) are not different enough to fully explain what is being observed, it rises even more intriguing question: why the discovery distances of Oort spike comets are statistically larger? 

An obvious answer might be, that the Oort spike comets are generally brighter than other LPCs at the same geocentric distance. It should be stressed, that we are not speaking here about their absolute brightness but their observed brightness at a large distance when they start their observable activity and therefore are discovered. 
This in turn does not seem to be an observational bias but a consequence of their different physical properties. Furthermore a number ratio between large- and small-perihelion LPCs seems to change  gradually with their $1/a_{\rm ori}$ in a significantly wider range of $1/a_{\rm ori}$ than the $1/a$-range of the Oort spike (see section~\ref{sec:a_original}). Today it is difficult to decide what is the main reason for this effect. One of the interesting
interpretations can be that this is a result of different processes responsible for cometary activity at different distances from the Sun and a gradually forming crust on the surface of the cometary nucleus in subsequent perihelion passages near the Sun. Another, or additional, supposition might be that dynamically new LPCs \citep[that is LPCs with previous perihelia outside the inner part of our planetary system, see][]{dyb-hist:2001,kroli_dyb:2017} are on average larger than dynamically old LPCs and during the first few passages near the Sun, their sizes are significantly reduced. Recent WISE/NEOWISE discoveries \citep{Bauer:2015,Bauer:2017} suggest that the LPCs are on average almost twice as large as short period comets. So maybe dynamically new comets are, on average, even bigger. Unfortunately, the dynamical status is known only for about 50~per~cent of Oort spike comets discovered so far. We can estimate that only $\sim$100 objects within the Oort spike comets are dynamically new objects, however, we can not identify them all yet. Our project on obtaining the dynamical status of all Oort spike comets, started many years ago, is still ongoing but it is a very time consuming task. This forced us to investigate the sample of Oort spike comets instead of the sample of dynamically new comets, although the latter would be more interesting from a dynamical point of view and will be done in future.   

\vspace{0.1cm}

\subsection{Discovery elongations versus discovery distances}\label{sec:elongation}

It is obvious that seeing a dim and diffuse object in small elongations is much more difficult than in large ones. And this is clearly visible in the case of LPCs, see second and third part of Table~\ref{table:discovery_distances}. One can see that the significant differences in statistics start from $\Delta_{\rm disc}$> 5\,au. These cometary detections since 1970 show that when we look at the sky in large elongations, we can easily discover a comet over 5\,au from the Earth. Statistics show that more Oort spike comets were discovered at $\Delta_{\rm disc}$>5\,au than on smaller $\Delta_{\rm disc}$. In a group of remaining LPCs with definitive orbits we have only about 30~per cent of objects having large $\Delta_{\rm disc}$. Thus, in elongations greater than 90\degr ~we can have a greater chance to observe quite a different type of cometary activity than the standard water sublimation starting around 3-5\,au from the Sun.

In the opposite case of small elongations, it seems that we have a serious problem with the detection of a comet beeing further than 5\,au from the Earth at the moment of observation, because statistics show here two and four times less LPCs  having larger $\Delta_{\rm disc}$ for Oort spike comets and remaining LPCs, respectively.

Therefore, a sample of LPCs discovered at large elongations seems to be more suitable for discussing the distribution of comets in a function of $\Delta_{\rm disc}$ as giving a more complete sample to larger distances from the Earth and the Sun (it is worth to notice that in such a geometry the heliocentric distance at the moment of discovery is always larger than $\Delta_{\rm disc}$) than the sample of LPCs discovered in small elongations. Additionally, this statistics limited to large elongations, though similar to statistics based on the full sample of LPCs (compare with general statistics of elongations presented in the first part of Table~\ref{table:discovery_distances}), is significantly less sensitive to the existence of a group of parabolic comets. One can see that for  large discovery elongations we have only 16~parabolic comets distributed almost uniformly in the range 0< $\Delta_{\rm disc}$<10\,au, whereas in the case of full statistics we have as many as 29~parabolic comets in the narrower range of 0< $\Delta_{\rm disc}$<2\,au, i.e more than Oort spike comets in the same range.

Conclusion from the sample limited to large elongations is that the number ratio of Oort spike comets and remaining LPCs sharply increases in function of $\Delta_{\rm disc}$ and is about 1.6 for $\Delta_{\rm disc}$>5\,au (column~5 in the second part of Table~\ref{table:discovery_distances})

\subsection{Perihelion distances}\label{sec:perihelion}

\begin{figure}
\includegraphics[width=8.9cm]{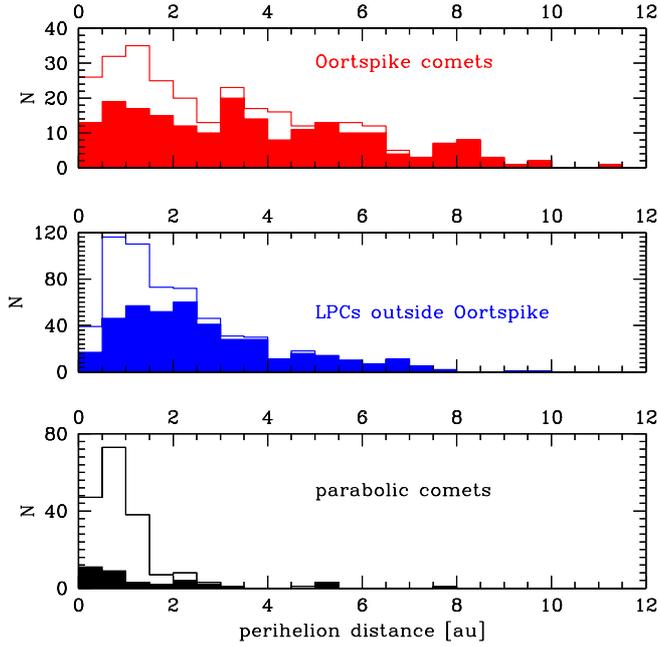} 
\caption{Distribution of Oort spike (upper red histogram), parabolic (lowest black histogram) and remaining LPCs (middle blue histogram) in function of perihelion distances. Fully coloured histograms describe LPCs discovered since 1980 whereas the distributions for LPCs discovered since 1801 are shown by coloured lines.}\label{fig:perihelion_distance}
\end{figure}

The differences in $\Delta_{\rm disc}$  for Oort spike comets and remaining LPCs discussed in previous two paragraphs obviously translate into the differences of perihelion distance distributions for these samples of LPCs. 
Fig.~\ref{fig:perihelion_distance} shows histograms for Oort spike comets (upper panel), remaining LPCs (middle panel) and parabolic comets (lowest panel) for objects discovered in the years 1980--2017 (full red, blue and black histograms) and in the 1801--2017 period  (outlined red, blue and black histograms, respectively).
As we can expect from the previous discussion, the distribution of Oort spike comets drops much slower (upper red histograms in Fig.~\ref{fig:perihelion_distance}) than for remaining LPCs (middle blue histograms) as $q$ increases.
A similar trend for a wider distribution of Oort spike comets was also described by \citet[][see his Fig.~2]{neslusan:2007} when he compared Oort spike comets (he called them dynamically new comets, however as we mention above, this term is not appropriate for comets defined solely by the condition: $1/a_{\rm ori}<10^{-4}$\,au$^{-1}$).   His work was based on \cite{marsden-cat:2003}. 

Generally, we notice here that all three distributions describing the samples of comets discovered since 1801 are significantly more peaked in the small perihelion distances than the respective distributions describing the discoveries in the 1980--2017 period. This is the result of a significant increase in observation sensitivity during the last two hundred years. Thanks to a rich sample of LPCs analysed here, we can next exclude comets discovered before 1980, which does not make the sample substantially smaller, and gives a more interesting comparison due to a more homogeneous sample in terms of observation capabilities.

We focus here on distributions based on discoveries after 1980 also because in this period we only have a few parabolic comets that could have only a minor impact on the  distributions of LPCs. One can notice that the distribution of Oort spike comets is only slightly decreasing (in a rough approximation it is almost uniform) in the range 0--6\,au for comets discovered in this period (full red histogram), and starting from $q=6$\,au a systematic decrease is observed up to 10\,au. 
Distribution of the remaining LPCs is different (full blue histogram). It increases between 0-1.5\,au (first three bins), peaks around 1.5-2.5\,au (next two bins) and rapidly decreases between 2.7-7\,au; only a few comets with $q>7.0$\,au were discovered.

Therefore, the conclusion is that the observed perihelion distance distributions of all three samples of LPCs correspond with our result drawn above that the Oort spike comets are discovered statistically at larger distances so such comets with large perihelion distances are more numerous. 

\section{Distribution of 1/\lowercase{a}-original for LPC\lowercase{s} discovered in the period 1801-2017}\label{sec:a_original}

\begin{figure}
\includegraphics[width=8.9cm]{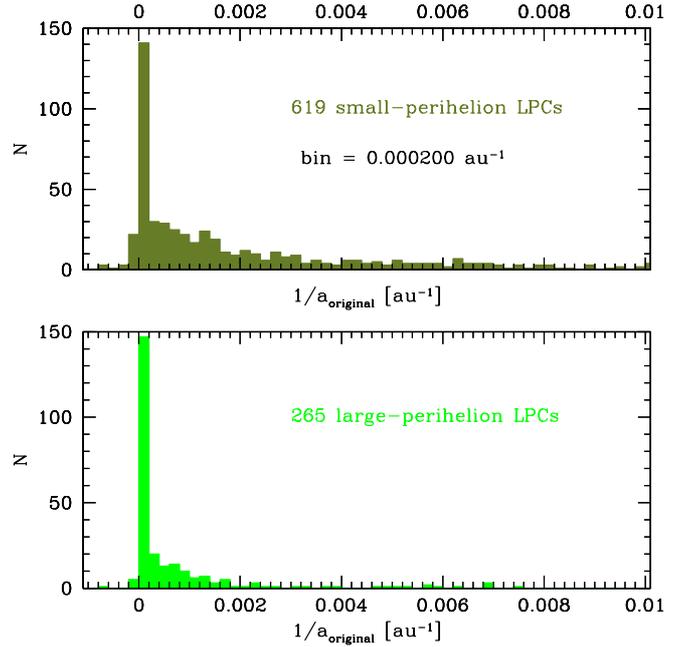} 
\protect\caption{\label{fig:qsmall_qlarge} Distribution of small-perihelion LPCs (upper panel) and large-perihelion LPCs (lower panel) in function of $1/a$-original for the bin width of 0.0002\,au$^{-1}$. Horizontal scale covers  the range of $1/a$-original for LPCs having orbital periods $\ge$ 1000\,yr.}
\end{figure}

\begin{center}
	\begin{table}
		\caption{Ratio of large-perihelion number of comets to small-perihelion comets in function of $1/a_{\rm ori}$ for LPCs with semimajor axis larger than 100\,au. Principal numbers are given for LPCs discovered after 1980, whereas statistics for LPCs detected since 1801 are given in parantheses; see text for details.} \label{table:ratio_large_small_in_a_bins}
		\centering
\setlength{\tabcolsep}{1.2pt} 	
		\begin{tabular}{cccc}
\hline
  Range of $1/a_{\rm ori}$ & \multicolumn{2}{c}{Number of LPCs}  & Ratio of large/small     \\
  $[$in au$^{-1}]$         & $q<3.1$\,au & $q>3.1$\,au           & perihelion LPCs   \\
\hline 
			&&& \\
			        $<$ 0.000000 &  23 (34)  &  4  (6)   & 0.17 (0.18)    \\  
			0.000000 -- 0.000040 &  38 (62)  &  46 (55)  & 1.21 (0.89)    \\  
			0.000040 -- 0.000100 &  32 (62)  &  59 (68)  & 1.84 (1.10)    \\
			0.000100 -- 0.000200 &  ~7 (16)  &  22 (22)  & 3.14 (1.37)    \\  
			0.000200 -- 0.000500 &  21  (39) &  29  (31) & 1.38 (0.79)    \\
			0.000500 -- 0.001000 &  45  (67) &  26  (28) & 0.58 (0.42)    \\
			0.001000 -- 0.010000 & 150 (244) &  40  (44) & 0.27 (0.18)    \\
			&&& \\
			        $<$ 0.000200 & 100 (174) & 131 (151) & 1.31 (0.87)    \\  
			\hline
		\end{tabular}
	\end{table}
\end{center}

For near-parabolic orbits the most indeterminable parameter is 1/a. While we might use the remaining elements as more or less reliable approximations we can state nothing on the shape of these orbits, except that their eccentricities are probably greater than 0.97 or so, see the examples presented in Table~\ref{table:parabolic}. For this reason in the present and the next sections we discuss the statistics and distributions excluding parabolic comets, however, we try to take them into account in some general conclusions drawn at the end of this section. 

Distributions of comets having small and large perihelion distances in a function of a semimajor axis of their original barycentric orbits are presented in Fig.~\ref{fig:qsmall_qlarge}. A qualitatively similar difference in distributions was observed by \citet[][see his Fig.~1]{neslusan:2007} for the subgroup of LPCs with $q<1$\,au and the full sample of LPCs discovered up to 2003.

\noindent The significant difference between two distributions of large- and small-perihelion LPCs presented here is easy to see. The large-perihelion LPCs distribution (lower green histogram) decreases much faster towards the shorter semimajor axes than the distribution of small-perihelion LPCs (upper olive histogram). This, of course, translates into a variable number ratio between these two subsamples of LPCs in each of these bins. In the second bin from the left (semimajor axis longer than 5000\,au) we have  a little bit greater number of large-perihelion LPCs  than small-perihelion LPCs but the whole subsample of small-perihelion LPCs is 2.3 times more numerous.

\begin{figure}
\includegraphics[width=8.9cm]{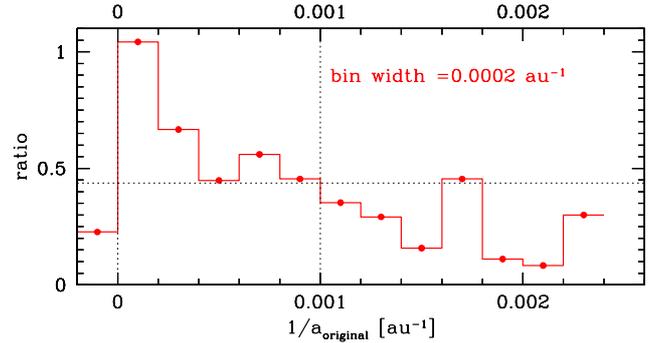} 
\protect\caption{\label{fig:qsmall_qlarge_ratio} Number ratio  of large-perihelion LPCs to small-perihelion LPCs in function of $1/a$-original for the same bins as in Fig.~\ref{fig:qsmall_qlarge}, but only for the range of the first thirteen of them ($1/a_{\rm ori}<0.0024$\,au$^{-1}$). Horizontal dotted black line indicates the mean number ratio of large-perihelion to small perihelion LPCs averaged over the entire sample of LPCs with $e\neq 1$.}
\end{figure}

In the next few bins a tendency to more or less systematic decrease in the contribution of large-perihelion LPCs relative to small-perihelion comets for a semimajor axis  shorter than 5000\,au is well visible in    Fig.~\ref{fig:qsmall_qlarge_ratio}.
Generally, for $a_{\rm ori}>1000$\,au the number ratio is above the horizontal dotted line representing the mean number ratio of large-perihelion comets to small-perihelion objects averaged over the entire sample of LPCs. Towards the smaller original semimajor axis the tendency is reversed and the number ratio in consecutive bins is generally smaller than the mean number ratio except for the interval of 0.0016\,au$^{-1}\,<1/a_{\rm ori}<0.0018$\,au$^{-1}$ (555\,au\,$<a_{\rm ori}<625$\,au). However for $a_{\rm ori}<1000$\,au we have rather poor statistics, especially for large-perihelion LPCs. 

Table~\ref{table:ratio_large_small_in_a_bins} shows the number ratio of large-perihelion to small-perihelion LPCs divided into seven different ranges of $1/a_{\rm ori}$, where the first three rows cover the Oort spike and in the next four rows are LPCs with more tight orbits. To reduce the impact of parabolic comet, only LPCs discovered after 1980 are taken into account. Statistics are poor only for $1/a_{\rm ori} <0$ and for the range $0.000100$\,au$^{-1} < 1/a_{\rm ori}< 0.000200$\,au$^{-1}$.  However, similar behaviour is also observed for all LPCs discovered after 1800, see numbers given in Table~\ref{table:ratio_large_small_in_a_bins} in parantheses.


It is clearly visible that the contribution of large-perihelion comets systematically increases(for $1/a_{\rm ori}<0.000200$\,au$^{-1}$) and next decreases (for $1/a_{\rm ori}>0.000200$\,au$^{-1}$) along with the shortening of their original semi-major axes. This decrease in ratio for semimajor axis shorter than 5000\,au is evident and is not significantly contaminated by omitting parabolic comets in these statistics. It is worth noting that the ratio of large-perihelion to small-perihelion LPCs drops more than three times in this $1/a_{\rm ori}$-interval (from 1.4 to 0.3, see Table~\ref{table:ratio_large_small_in_a_bins}).

Situation is different, however, with the initial percentage increase of large-perihelion LPCs within the Oort spike and a bit further ($1/a_{\rm ori}<0.000200$\,au$^{-1}$). It is necessary to take here into account the discoveries of parabolic comets  as well as the potential impact of NG~effects on this statistics. The latter is closely related to the existence of comets with formally hyperbolic orbits. It is worth to remember that orbits of these comets are only slightly hyperbolic and uncertainties of $1/a_{\rm ori}$ allow us to include these objects to the Oort spike comets with closed orbits. Additionally, all these orbits were determined ignoring NGF, and it should be reminded here that for individual comet with detectable non-gravitational effects, the orbit becomes generally more tight than purely gravitational orbit. Thus, we can assume that comets with a negative $1/a_{\rm ori}$ (85 percent of them are small-perihelion comets) could feed the other $1/a_{\rm ori}$-ranges if non-gravitational orbits could be determined. This would result in some decrease in the ratio value (last column of Table~\ref{table:ratio_large_small_in_a_bins}) for ranges within the Oort spike or even for $1/a_{\rm ori}<0.000200$\,au$^{-1}$. 
We can expect an additional reduction in this ratio after including parabolic comet to this statistics because as many as 34~parabolic comets have been discovered since 1980, and 29 of them have perihelia closer than 3.1\,au to the Sun. Thus, these comets reduce the percentage contribution of large-perihelion LPCs in the range of $1/a_{\rm ori}<0.000200$\,au$^{-1}$ from 57 to 51 per cent.  Unfortunately, it is not possible to prove whether the observed increase in ratio  within three ranges of positive semimajor axis up to $1/a_{\rm ori}=0.000200$\,au$^{-1}$  would still be then preserved.

\section{Observed shape of the Oort spike}\label{sec:Oortspike}

\begin{figure}
\includegraphics[width=8.9cm]{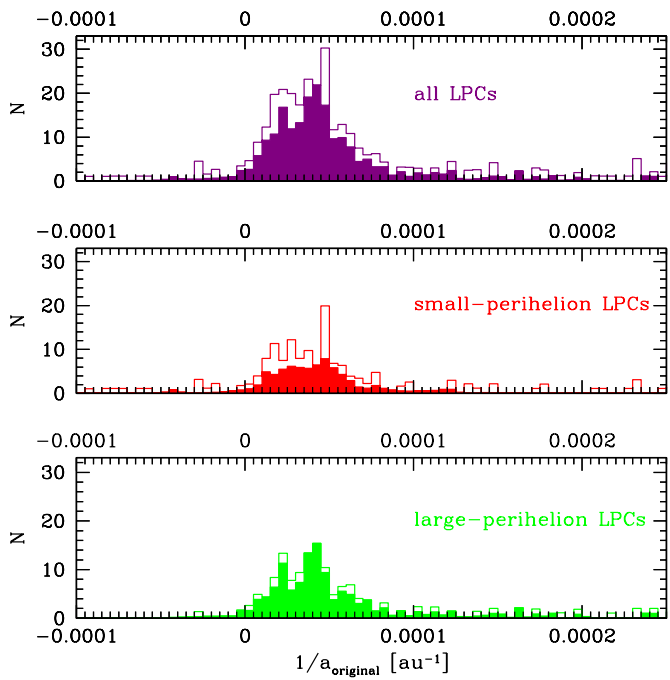} 
\caption{Distribution of $1/a$-original for LPCs with semimajor axes greater than 4\,000\,au. Narrow bins of width of 0.000005\,au$^{-1}$ were applied here. See text for a detailed explanation.}\label{fig:qsmall_qlarge_VCs_nominals}
\end{figure}

One third of all LPCs discovered between 1801--2017 with $e \ne 1$ creates the Oort spike (in a wider sense, i.e. with $1/a_{\rm ori} < 0.0001$\,au$^{-1}$, see Table~\ref{table:statistics}) and as much as 45 per cent of them are large-perihelion  comets ($q>3.1$\,au). This sample of Oort spike comets were investigated by us in detail in our previous papers (see KD17 and references therein). 
In KD17 we obtained the Oort spike shape taking into account the uncertainties of derived values of the original semimajor axes for 100~large-perihelion comets. It means that the Oort spike was constructed using the full swarms of 5001 VCs (virtual comets, including nominal orbit, see KD17 for a detailed description of this technique). To construct analogous histograms based on a complete sample of 287~Oort spike comets and other LPCs, we use a mixed technique here. Currently, we have original orbital solutions for 134~large-perihelion LPCs (KD17, and Kr{\'o}likowska and Dybczy{\'n}ski in prep.) and 109 small-perihelion LPCs \citep{krolikowska:2014, krol-sit-et-al:2014, kroli-dyb:2016, Kroli-Dyb:2018b}; Kr{\'o}likowska et al., in prep.). We used normalized full swarms of $1/a_{\rm ori}$-distributions for all above comets. The resulted distributions are given as fully coloured histograms presented in Fig.~\ref{fig:qsmall_qlarge_VCs_nominals} for large-perihelion comets (lowest panel), small-perihelion comets (middle panel) and all these LPCs together (upper panel). Next, we added nominal $1/a_{\rm ori}$-values for the remaining LPCs, and the cumulative result is presented with the contoured histograms in the same colour as the full histogram in Fig.~\ref{fig:qsmall_qlarge_VCs_nominals}. 

Let us stress that the  $1/a_{\rm ori}$-distribution of large-perihelion comets takes into account the semimajor axis uncertainties for 85~per cent of all observed Oort spike comets in this sample (lowest panel in Fig.~\ref{fig:qsmall_qlarge_VCs_nominals}). Thus, in this case the observed shape of the Oort spike is constructed quite precisely. Two local maxima around 20 and 40 in units of $10^{-6}$\,au$^{-1}$ are visible in this case. For a more detailed discussion of the possible source of this feature see KD17. 

It is much worse with the precision of the Oort spike reconstruction for small-perihelion comets. Oort spike shape reveals a fairly smooth maximum between 10 and 60 in units of $10^{-6}$\,au$^{-1}$ for comets with the detailed uncertainty statistics of $1/a_{\rm ori}$ (full histogram). This maximum can be only slightly extended due to the individual widths of $1/a_{\rm ori}$ swarms because, so far, we have been dealing with comets with orbits of the first quality class (with a typical uncertainty of $1/a_{\rm ori}$-value below $2\times 10^{-6}$\,au$^{-1}$), rarely of the second class. 

Only a set of 25 small-perihelion 'parabolic' comets in JPL database  and three other LPCs (C/1904~Y1, C/1940~S1, and C/1959~Y1) have poor quality orbits ($1/a_{\rm ori}$-uncertainty greater than the width of Oort spike, for examples see Table~\ref{table:parabolic}). Additionally, the maxima of $1/a_{\rm ori}$ distributions of VCs swarms for almost all these comets are located far beyond the right border of Fig.~\ref{fig:qsmall_qlarge_VCs_nominals}, except five comets (C/1904~Y1, C/1925~X1, C/1927~A1, C/1959~Y1, and C/2007~Q1) with a nominal orbit within the Oort spike. We have checked that the cumulative $1/a_{\rm ori}$-distribution for all these comets creates only a barely elevated background in Fig.~\ref{fig:qsmall_qlarge_VCs_nominals}, fluctuating in the range of about 0.02--0.05 LPCs/per~bin within the Oort spike. Thus, these comets are not responsible for such a quite smooth and rather wide Oort spike maximum for small-perihelion LPCs (red full histogram).

\noindent However, the precise distribution using full swarms of VCs is constructed here only for 50~per cent of known small-perihelion Oort spike comets. Among the remaining small-perihelion Oort spike comets (that contribute to the outlined distribution with their nominal orbit only) there is a larger fraction of second class orbits. Thus, we can expect that their individual $1/a_{\rm ori}$-distributions would be more diffuse and the shape of the Oort spike for small-perihelion comets fully based on individual $1/a_{\rm ori}$-distributions of VCs may turn out to be similarly wide as presented now fully coloured distribution, or (which is less likely) indicate traces of  two narrower maxima.

\section{Summary}\label{sec:conclusions}

This paper presents a rich picture of statistical behaviour of a complete set of 1065 LPCs discovered since 1800. After a careful discrimination between hyperbolic, parabolic and so called 'other' comets (i.e. having the definitive eccentricity smaller than one) on the basis of their osculating orbit elements
 as well their original semimajor axis we construct two samples of LPCs: Oort spike comets and LPCs with more
tight orbits. Next,  we show several interesting statistics of their discovery circumstances and their orbital parameters.

We discuss in detail the influence of a group of parabolic comets on the presented statistics. To  minimize their impact on a various kind of LPCs statistics, we determined and used the definitive orbits for about 40~parabolic comets from JPL database, where more than 20\, of these orbits were determined in the present study (see Table~\ref{table:parabolic}). Next, we show that this reduced subset of 181~LPCs with indeterminable eccentricities, and semimajor axes, do not change our statistics in a significant manner. 

We confirmed several well-known properties of the observed LPCs population and show how they evolve with time, which means that also  with the increase of observational capabilities. Additionally, we noticed one new and very expressive feature: the percentage of large-perihelion comets is significantly higher in a group of the Oort spike objects than for LPCs moving on more tight orbits. This percentage grows continuously from the half of the previous century and  it is not clear whether it has already reached its maximum or will continue to grow (see  Fig.~\ref{fig:ratio} and Table~\ref{table:ratio_large_small}).

To explain this feature we studied in detail discovery circumstances for all studied LPCs, and found that the Oort spike comets are discovered statistically at larger geocentric and heliocentric distances, especially in the last decades. This effect can be a direct consequence of a well-known comet fading process in subsequent apparitions due to their surface ageing. This interpretation is strongly supported by presented continuous decrease of the ratio of large- to small-perihelion LPCs with the increasing value of their $1/a_{\rm ori}$-values in a quite wide interval of at least  0 $<1/a_{\rm ori}<$ 0.001\,au$^{-1}$ (see Fig.~\ref{fig:qsmall_qlarge_ratio} and Table~\ref{table:ratio_large_small_in_a_bins}. However, the second possibility, not excluding the first, is that this may also be the effect of some differences in actual perihelion distance distributions between Oort spike comets and LPCs moving on more tight orbits, especially in the range of $q>4$\,au (see Fig.~\ref{fig:perihelion_distance}).

We also noticed that a subgroup of LPCs discovered at elongations $>90\degr$ is much more differentiated both in dynamical and physical LPCs properties, mainly due to the lack of most observational selection effects.

The results shown in this paper are based on the division of all LPCs with known eccentricities into two subgroups depending on their $1/a_{\rm ori}$-values: Oort spike comets ($1/a_{\rm ori}<10^{-4}$\,au${-1}$) and remaining LPCs. It would be much more interesting to additionally divide the Oort spike comets into dynamically new and dynamically old objects, and to investigate differences in these three constructed subgroups of LPCs. For this, however, it is necessary to know previous perihelion distances for these objects, see for example \citet{dyb-hist:2001} or KD17. So far, we have obtained such the data for only a half of small-perihelion Oort spike comets, and about 85~per~cent of large-perihelion Oort spike comets (see sec.~\ref{sec:Oortspike}). We are still working in a long-standing project focusing on the study of past and future dynamical evolution of Oort spike comets. This task is significantly more difficult for small-perihelion comets,  where non-gravitational effects are important and have to be included in the process of osculating orbit determination. This work is in progress.



\section*{Acknowledgements}

We would like to thank Maik Meyer for making his Catalogue of Cometary Discoveries available to us and an anonymous Referee for comments on interpretation the observational findings.  
This research was partially supported by the project 2015/17/B/ST9/01790 founded by the National Science Centre in Poland.

\bibliographystyle{mnras}

\bibliography{moja24.bib}


\end{document}